\documentclass{emulateapj}
\usepackage{epsfig}
\usepackage{graphics}
\usepackage{epstopdf}
\newlength{\colwidth}
\newcommand{\HI}{\ion{H}{1}}

\newcommand{\CIII}{\ion{C}{3}}
 
\newcommand{\CIV}{\ion{C}{4}} 
\newcommand{\SiIII}{\ion{Si}{3}} 
 
\newcommand{\SiIV}{\ion{Si}{4}} 
 
\newcommand{\OVI}{\ion{O}{6}} 
\newcommand{\OV}{\ion{O}{5}} 

\newcommand{\tlya}{\tau_{\rm Ly{\alpha}}}
\newcommand{\tovi}{\tau_{\rm OVI}}
\newcommand{\tsiiv}{\tau_{\rm SiIV}}
\newcommand{\tciv}{\tau_{\rm CIV}}

\newcommand{\thi}{\tau_{\rm HI}}

\newcommand{\ttru}{\tau_{\rm true}}

\newcommand{\tmin}{\tau_{\rm min}}

\newcommand{\lya}{Ly$\alpha$} 
\newcommand{\lyb}{Ly$\beta$}

\lefthead{Aguirre et al.}
\righthead{Oxygen in the IGM}

\def\gsim{\;\rlap{\lower 2.5pt
 \hbox{$\sim$}}\raise 1.5pt\hbox{$>$}\;}
\def\lsim{\;\rlap{\lower 2.5pt
   \hbox{$\sim$}}\raise 1.5pt\hbox{$<$}\;}

\def\kms{\rm\,km\,s^{-1}}

\def\mpc{{\rm\,Mpc}}

\def\kms{{\rm\,km\,s^{-1}}}
\def\K{{\rm\,K}}
\def\spose#1{\hbox to 0pt{#1\hss}}
\def\lta{\mathrel{\spose{\lower 3pt\hbox{$\mathchar''218$}}
     \raise 2.0pt\hbox{$\mathchar''13C$}}}
\def\gta{\mathrel{\spose{\lower 3pt\hbox{$\mathchar''218$}}
     \raise 2.0pt\hbox{$\mathchar''13E$}}}

\shorttitle{Oxygen in the IGM}
\shortauthors{Aguirre et al.}

\begin{document}
	
\title{Metallicity of the intergalactic medium using pixel
statistics:\\ IV. Oxygen.\altaffilmark{1}} 
\altaffiltext{1}{Based on
public data obtained from the ESP archive of observations from the UVES
spectrograph at the VLT, Paranal, Chile and on data obtained at the W. M. Keck 
Observatory, which is operated as a scientific partnership among the
California Institute of Technology, the University of California, and
the National Aeronautics and Space Administration. The W. M. Keck
Observatory was made possible by the generous financial support of the
W. M. Keck Foundation.}

\author{ Anthony Aguirre\altaffilmark{2},
Corey Dow-Hygelund\altaffilmark{2},
Joop~Schaye\altaffilmark{3},
Tom~Theuns\altaffilmark{4}
}
\altaffiltext{2}{Department of Physics, University of California at
  Santa Cruz, 1156 High Street, Santa Cruz, CA  95064; aguirre@scipp.ucsc.edu,
  godelstheory@gmail.com} 
\altaffiltext{3}{Leiden Observatory, Leiden University, P.O. Box 9513,
  2300 RA Leiden, The Netherlands; schaye@strw.leidenuniv.nl} 
  \altaffiltext{4}{Institute for Computational Cosmology, University of Durham, South Road, Durham DH1 3LE; tom.theuns@durham.ac.uk} 

\setcounter{footnote}{0}

\begin{abstract}

We have studied the abundance of oxygen in the IGM by analyzing
\OVI, \CIV, \SiIV, and \HI\ pixel optical depths derived from a set of 
high-quality VLT and Keck spectra of 17 QSOs at $2.1\la z \la
3.6$. Comparing ratios  
$\tovi/\tciv(\tciv)$ to those in realistic, synthetic spectra drawn
from a hydrodynamical simulation and comparing to existing constraints on [Si/C] places strong constraints on the ultraviolet background (UVB) model using weak priors on allowed values of [Si/O]: for example,  a quasar-only background yields [Si/O] $\approx 1.4$, highly inconsistent with the [Si/O] $\approx 0$ expected from nucleosynthetic yields and with observations of metal-poor stars. 
Assuming a fiducial quasar+galaxy UVB consistent with these constraints yields a primary
result that [O/C] = 0.66 $\pm$ 0.06 $\pm$ 0.2; this result is sensitive to gas with overdensity $\delta \gtrsim  2$. Consistent results are obtained by similarly comparing
$\tovi/\thi(\thi)$ and $\tovi/\tsiiv(\tsiiv)$ to simulation values, and also
by directly ionization-correcting $\tovi/\thi$ as function of $\thi$
into [O/H] as a function of density.
Subdividing the sample reveals no evidence for evolution, but low- and
high-$\thi$ samples are inconsistent,  
suggesting either density-dependence of [O/C] or -- more likely --
prevalence of collisionally-ionized gas at high density.
\end{abstract}

\section{Introduction}
\label{sec-intro}

The enrichment of the intergalactic medium (IGM) with heavy elements has, over the past decade, become a key tool in understanding star and galaxy formation by providing a fossil record of metal formation and galactic feedback.  

Absorption line spectroscopy has revealed, among other findings, that
the low-density ($\delta \equiv \rho/\left <\rho\right > \la 10$)
intergalactic medium (IGM), as probed by the \lya\ forest and through
\CIII, \CIV, \SiIII, \SiIV, \OV, \OVI, and other transitions, is at
least partly enriched at all redshifts and densities probed.  In
particular, recent studies indicate that: 
\begin{itemize}
\item When smoothed over large ($\sim 10 - 10^2$~kpc) scales, the
  abundance of carbon decreases as gas overdensity 
  $\delta$ does and has a scatter of $\sim 1\,$dex at fixed
  density. There is carbon in at least some gas at all densities down
  to at least the mean cosmic density, with the median carbon
  metallicity obeying [C/H] $\approx -3.5 + 0.65(\log \delta
  -0.5)$ at $z\approx 3$~\citep[Schaye et al.\ 2003, hereafter ][]{paper2}. 
\item On smaller ($\la 1$~kpc) scales the distribution of metals is
  less well known, but observations suggest that the metals may be
 concentrated in small, high-metallicity patches \citep{schaye2007}.
 \item There is no evidence for metallicity evolution 
  from redshift $z\approx 4$ to $z\approx 2$~\citep{paper2} and metals exist
  at some level at 
  $z \approx 5 -
  6$~\citep{2001ApJ...561L.153S,2003ApJ...594..695P,2006MNRAS.371L..78R,2006ApJ...653..977S}.  
\end{itemize}

In connection with this observed widespread distribution of metals, a
general picture has emerged that galactic winds -- driven largely from
young and/or starburst galaxies -- have enriched the IGM. The same
feedback may account for the dearth of low-luminosity galaxies
relative to the halo mass
function~\citep[e.g.,][]{1993MNRAS.264..201K,1999MNRAS.310.1087S,2003MNRAS.339..312S},
and also for the mass-metallicity relationship of
galaxies~\citep[e.g.,][]{2004ApJ...613..898T,2006ApJ...644..813E}.
However, a detailed understanding of the various feedback processes is
lacking and there are still open questions and controversies
concerning the time and relative importance of the various enrichment
processes, and concerning the implications for galaxy formation. 

Both theoretical modeling and observations of intergalactic (IG)
enrichment are now advancing to the point where comparison between the
two can provide crucial insight into these issues, but this comparison
is not without problems. 
Two key difficulties concern the ionization correction required
to convert observed ionic abundances into elemental abundances.  First, while
the oft-studied ions \CIV\, and \SiIV\, are observationally
convenient, they are poor probes of hot ($> 10^5\,$K) gas, because
the ion factions \CIV/C and \SiIV/Si both fall dramatically with
temperature.  Thus, the 
hot remnants of fast outflows might be largely invisible in these
ions.  Second, the dominant uncertainty in both the absolute and
relative abundance inferences stems from uncertainty in the spectral
shape of the ultraviolet ionizing background radiation (UVB). 
 
Analysis of oxygen, as probed by \OVI, has the potential to shed light
on both problems: this ionization state becomes prevalent in some of
the very phases in which  \CIV\, and \SiIV\, become rare, and its
abundance depends on the UVB shape differently than those of other
ions, helping break the degeneracy between abundances and UVB shape.
The challenge posed by \OVI\, is that at $z\gtrsim 2$ it is strongly
contaminated by both \lya\, and Ly$\beta$ lines, making its
identification and quantification difficult. Previous studies of high-$z$
oxygen enrichment using line
fitting~\citep{Carswell2002,Bergeron2002,Simcoe2004} or pixel 
statistics~\citep{2000ApJ...541L...1S,Telfer2002,Pieri2004} have
reliably detected oxygen in the IGM, and quantified its abundance in
relatively dense gas, but have not assessed the oxygen abundance with a
very large data sample, at very low-densities, or in a unified
treatment with other available ions.   

Here we extend to \OVI\, our application of the ``pixel optical depth"
technique~\cite[e.g., Aguirre, Schaye \& Theuns, hereafter][]{paper1}
to a large set of high quality VLT/UVES and Keck/HIRES spectra.  The
results, when combined with previous studies of \CIV\, and
\CIII~\citep{paper2} and of \SiIV\, and \SiIII~\citep[Aguirre
  et. al. 2004; hereafter][]{paper3}, give a comprehensive
observational assessment of IG enrichment by carbon, silicon and
oxygen, with significantly reduced uncertainties due to the UVB shape,
as well as new data on the importance of hot, collisionally ionized gas. 

We have organized this paper as follows.  In \S\S\ref{sec-data}
and~\ref{sec-overview} we briefly describe our sample of QSO
spectra. The analysis method is described briefly in
\S\ref{sec-overview} and then in greater depth in the remainder of
\S\ref{sec-meth}, with heavy reference to Papers I, II and III.
The basic results are given in \S\ref{sec-resrel} and discussed in
\S\ref{sec-discuss}. Finally, we conclude in \S\ref{sec-conc}. 

All abundances are given by
number relative to hydrogen, and solar abundance are taken to be
$({\rm O/H})_\odot = -3.13$, $({\rm C/H})_\odot = -3.45$, and  
$({\rm Si/H})_\odot = -4.45$
\citep{1989GeCoA..53..197A}.

\section{Observations}
\label{sec-data}

We analyze 17 of the 19 high-quality ($6.6\,\kms$ velocity resolution,
S/N $> 40$) absorption spectra of quasars used in Papers II and III.
The two highest-redshift spectra used in those previous studies were
excluded here because the severe contamination of the \OVI\, region by
\HI\ lines makes detection of \OVI\, nearly impossible and also
introduces very large continuum fitting errors in the \OVI\, region.
Fourteen spectra were taken with the UV-Visual Echelle
Spectrograph~\citep[UVES,][]{2000SPIE.4005..121D} on the Very Large
Telescope and three were taken with the High Resolution Echelle
Spectrograph~\citep[HIRES,][]{1994SPIE.2198..362V} on the Keck
telescope.  For convenience, the observed QSOs are listed in
Table~\ref{tbl:sample}. 
  
\begin{deluxetable}{llcccclll}
\tablecolumns{8}
\tablewidth{0pc}
\tablecaption{Observed quasars \label{tbl:sample}}
\tablecomments{Columns 1 and 2 contain the quasar name and corresponding Ly$\alpha$ emission redshift. Columns 3 and 4 contain the
minimum and maximum absorption redshifts considered and Column 5 contains the corresponding minimum observed wavelength.  The last column contains an estimate of the percentage uncertainty in the continuum fitting in the \OVI\ region.}
\tablehead{
\colhead{QSO} & \colhead{$z_{\rm em}$} & \colhead{$z_{\rm min}$} & \colhead{$z_{\rm max}$} & \colhead{$\lambda_{\rm
min}$ (\AA)} & \colhead{instrument} & \colhead{ref} & \colhead{Err}}
\startdata
Q1101-264   & 2.145 & 1.878 & 2.103 & 3050.00 & UVES & 1 & 1.6 \\  
Q0122-380   & 2.190 &  1.920 & 2.147 & 3062.00 & UVES & 2 & 0.6 \\  
J2233-606   & 2.238 &  1.963 & 2.195 & 3055.00 & UVES & 3 & 1.1 \\  
HE1122-1648 & 2.400 &  2.112 & 2.355 & 3055.00 & UVES & 1 & 1.4 \\  
Q0109-3518  & 2.406 &  2.117 & 2.361 & 3050.00 & UVES & 2 & 1.5 \\  
HE2217-2818 & 2.406 &  2.117 & 2.361 & 3050.00 & UVES & 3 & 1.6 \\  
Q0329-385   & 2.423 &  2.133 & 2.377 & 3062.00 & UVES & 2 & 1.2 \\  
HE1347-2457 & 2.534 &  2.234 & 2.487 & 3050.00 & UVES & 1,2& 2.5 \\  
PKS0329-255 & 2.685 &  2.373 & 2.636 & 3150.00 & UVES & 2 & 1.5 \\  
Q0002-422   & 2.76  &  2.441 & 2.710 & 3055.00 & UVES & 2 & 1.6 \\  
HE2347-4342 & 2.90  &  2.569 & 2.848 & 3428.00 & UVES & 2 & 1.5 \\  
Q1107+485   & 3.00  &  2.661 & 2.947 & 3644.36 & HIRES & 4 & 2.3 \\ 
Q0420-388   & 3.123 &  2.774 & 3.068 & 3760.00 & UVES & 2 & 1.8 \\  
Q1425+604   & 3.20  &  2.844 & 3.144 & 3736.20 & HIRES & 4 & 2.1 \\ 
Q2126-158   & 3.268 &  2.906 & 3.211 & 3400.00 & UVES & 2 & 2.0 \\  
Q1422+230   & 3.62  &  3.225 & 3.552 & 3645.24 & HIRES & 4 & 3.1 \\ 
Q0055-269   & 3.655 &  3.257 & 3.586 & 3423.00 & UVES & 1 & 4.0 
\enddata
\tablerefs{(1) Kim et al.\ 2002; (2) Kim et al.\ 2004; (3)
  Kim, Cristiani, \& D'Odorico 2001; (4) Rauch et al.\ 1997.}
\end{deluxetable}  

Regions within $\Delta v =
\max(4000,8\,\mpc\,H(z)/h)~\kms$ from the quasars, where $H(z)$ is the Hubble
parameter at 
redshift $z$ extrapolated from its present value ($H_0 \equiv
100h~\kms\,\mpc^{-1}$) assuming $(\Omega_m,\Omega_\Lambda) =
(0.3,0.7)$, were excluded to avoid proximity effects. Regions thought
  to be contaminated by absorption features that are not 
present in our simulated spectra (e.g., damped Ly$\alpha$ systems)
were also excluded from the analysis.

Lyman continuum contamination increases significantly towards lower
wavelengths, whereas (as described below) our correction for this contamination assumes that it is non-evolving.  To mitigate this effect, only the red portion [$\geq
{\rm med(\textit{z})}$] of the QSO spectra used in Papers II and III
is analyzed in this Paper. As in \cite{2000ApJ...541L...1S}, this was
found to result in smaller errors than using the full region. 

Further details concerning the sample and
data reduction are given in Paper II (\S2).

\section{Method}
\label{sec-meth}

The pixel optical depth method we use for measuring \OVI\ is similar to that described in Papers I, II, and III.  Section \ref{sec-overview}
contains a brief outline of the method; \S\ref{sec-confit} and
\S\ref{sec-recovery} describe continuum fitting and contamination
corrections, which have been changed slightly from the methods described in
Papers II and III; \S\ref{sec-ovitest} describes tests of the
recovery, and \S\ref{ioncorr} discusses the ionization balance of the
relevant species, and describes how ionization corrections are performed. 

\subsection{Overview}
\label{sec-overview}

The basic method for analysis of each QSO spectrum is as follows:

\begin{enumerate}

\item Optical depths due to \HI\
\lya\ ($\lambda1216$~\AA) absorption are recovered for all pixels in the
\lya\ forest region, using higher-order Lyman lines to estimate optical depths for saturated pixels.

\item The pixel
optical depth at the corresponding wavelengths of the metal
lines \OVI\ ($\lambda\lambda1032,1038$), \CIV\ ($\lambda\lambda1548,1551$), and \SiIV\ ($\lambda\lambda1394,1403$) are recovered, making several corrections to reduce contamination and noise.

\item The recovered optical depth in one transition is compared with that of another, by binning the pixels in terms of the optical depth of \HI, \CIV, or \SiIV, and plotting the median (or some other percentile of) optical depth of \OVI. A correlation then indicates a detection of \OVI\ absorption.  An example is shown in Fig.~\ref{fig:oviscat}. 

\begin{figure*}
\plotone{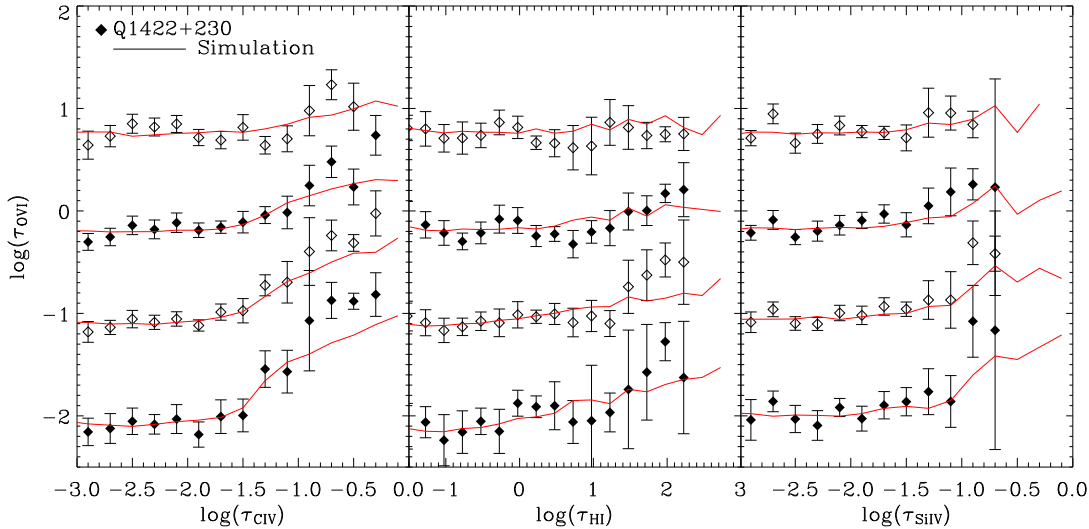} 
\figcaption[]{\OVI\ optical depths as a function of $\tciv$ (left), $\thi$ (middle), and $\tsiiv$ (right).  From bottom to top, the points represent the 31st, median, 69th, and 84th percentiles.  Data points are plotted with 1 $\sigma$ error bars.  The red lines are predictions from simulations with $\langle {\rm [C/H]}\rangle = -3.8 +
0.65 \delta$, $\sigma=0.70$, [Si/C]$=0.77$, [O/C]$=0.64$.
The observed and simulated 31st, 69th and 84th percentiles have been
been offset vertically by -0.5, 0.5 and 1.0 dex, respectively. 
\label{fig:oviscat}}
\end{figure*}

\end{enumerate}

As was done in Papers II and III, an identical analysis is applied to
synthetic spectra generated using a cosmological, hydrodynamical
simulation, kindly provided by Tom Theuns. For 
each observed quasar we generate 50 corresponding simulated spectra
with the same noise properties, wavelength coverage, instrumental
broadening, and pixel size as the observed spectra.  For each UVB
model (of which several are used; see below) the carbon distribution
as measured in Paper II, and the value of [Si/C] from Paper III, are
imposed on the fiducial spectra.  An oxygen abundance is assigned by
assuming a \textit{constant, uniform} value of [O/C].  Ionization
balances are calculated using CLOUDY\footnote{See 
\texttt{http://www.pa.uky.edu/$\sim$gary/cloudy}.}(version 94; see
Ferland et al.\ 1998 and Ferland 2000 for details).  A direct
comparison of the results from these simulated and observed
spectra allows for inferences about the distribution of oxygen,
carbon, and silicon.  The same simulation was used in Papers I (\S3),
II, and III, to which the reader is referred for details. 

This study employs the  identical UVB models used in Papers II (\S4.2)
and III, excluding model ``QGS3.2''.  All models are from Haardt \&
Madau (2001, hereafter HM01)\footnote{The 
data and a description of the input parameters can be found at
\texttt{http://pitto.mib.infn.it/$\sim$haardt/refmodel.html}.}.  These
have been
renormalized (by a redshift-dependent factor) such that the simulated
spectra match the observed evolution of the mean \HI\
Ly$\alpha$ absorption (Paper II). The fiducial model, ``QG'', includes
contributions from both galaxies (with a 10\% escape fraction for
ionizing photons) and quasars; ``Q'' includes only quasars; ``QGS'' is
an artificially softened version of QG: its flux has been reduced by a
factor of ten above 4~Ryd.  The UVB used in the simulation only affects the IGM temperature, and was chosen to match the measurements by~\citet{2000MNRAS.318..817S}.

\subsection{Continuum fitting}
\label{sec-confit}

A major source of error in \OVI\ optical depths is continuum fitting
in the \OVI\ absorption region, where contamination by \lya\ and \lyb\ lines is
heavy. To make this fitting as accurate as possible, and to furnish an
estimate of the continuum fitting error, we have applied the following
procedure to the region analyzed for \OVI\ absorption (in the case of
observed spectra, this was done after the spectra had been continuum
fitted by eye as described in
Paper II \S2): 
\begin{enumerate}
\item{The spectral region is divided into 20\,\AA\ (rest-frame) segments.}
\item{In segments with large unabsorbed regions, an automatic
  continuum fitting algorithm is applied (see \S5.1 of Paper II), in
  which pixels $> 1\sigma$ below the continuum are iteratively
  removed.} 
\item{In segments without large unabsorbed regions, we identify small
  unabsorbed 
regions or regions absorbed only in \lyb; the
latter are identified by superimposing the region of the spectrum
corresponding to  Ly$\alpha$ absorption. The continuum level of the segment
is fit by minimizing the deviation of identified unabsorbed regions
from unit flux, and deviation of the \lyb\ regions from the
corresponding scaled Ly$\alpha$ features.}
\item{A spline is interpolated between the fits to all segments and
  the spectrum is rescaled by this spline. 
}
\end{enumerate}
This procedure was applied to all observed spectra as well as to to one
simulated spectrum per observed spectrum, where a 10-20\% error
in the continuum was introduced on scales of 1, 4 and 16 segments. The
median absolute errors remaining after blindly fitting the continua of
the simulated spectra are given in Table \ref{tbl:sample} as an estimate of continuum
fitting errors in the corresponding observed spectra. Because the
procedure is not fully automatic, we were unable to apply it to all of
the simulated spectra.

The region redwards of \lya\ was fit in both simulated and observed
spectra using the procedure described in \S5.1 step I of Paper II.
Continuum fitting errors are much smaller for this region
($\sim$0.01$\%$).   

\subsection{Correcting for contamination}
\label{sec-recovery}

After continuum fitting the spectra, \HI\ (Ly$\alpha$) optical depths $\thi$ are derived for each pixel between the quasar's Ly$\alpha$ and Ly$\beta$ emission wavelengths, save for regions close to the quasar to avoid proximity effects
(see \S2). If Ly$\alpha$ is saturated (i.e.,
$F(\lambda) < 3\sigma(\lambda)$, where $F$ and $\sigma$ are the
flux and noise arrays, see Paper I, \S4.1; Paper II, \S5.1, step 2),
higher-order 
Lyman lines are used to estimate $\thi$.    

Corresponding \OVI, \SiIV, and \CIV\ optical depths ($\tovi$,
$\tsiiv$, $\tciv$) are subsequently derived for each \HI\ pixel.  We
exclude regions of the quasar spectrum that are contaminated by
absorption features that are not included in our simulated
spectra, such as \lya\ lines with damping wings.  For $\tsiiv$ and 
$\tciv$, corrections are made for self-contamination and contamination
by other metal lines, as described in Paper I, \S4.2. 

As shown in Fig.~\ref{fig:oviscat}, when plotting each percentile in $\tovi$ absorption against absorption in some other ion, the correlation disappears below some \OVI\
optical depth $\tau_{\rm min}$ (corresponding to a value $\tau_c$ in the other ion) that is determined by noise, continuum fitting errors, and contamination by other lines.
These effects may then be corrected for by subtracting $\tau_{\rm min}$ from the
binned optical depths, thus converting most points below $\tau_c$ into
upper limits. For each realization and for each percentile, we compute
$\tau_{\rm min}$ as the given percentile of optical depth for the set
of pixels with optical depth $< \tau_c$.  We use values $\tau_c=0.01$
when binning in \CIV\, or \SiIV, and of $\tau_c=0.1$ when binning in
\HI, as we never see a correlation extending below these
values.\footnote{In Paper II we used functional fits to determine
  $\tau_c$. For OVI the correlations are generally less 
strong than for CIV and we fix $\tau_c$ ``by hand.''}

The error on $\tau_{\rm min}$ for an individual realization is computed by dividing the spectrum into 5\,\AA\ segments, then bootstrap-resampling the spectrum by choosing these chunks with replacement, and finally computing the variance of $\tau_{\rm min}$ as computed from 100 such resampled spectra. When the realizations are combined, $\tau_{\rm
min}$ is instead computed as the median among the realizations, and the error
on this value is computed by bootstrap-resampling the realizations. 
For further details see Paper II, \S5.1, step 4 and Paper III, \S3.4.

As noted above, \lya\ and higher Lyman transitions heavily contaminate
the \OVI\ absorption regime.  This can add substantial error in the
recovered $\tovi$.  Two corrections are made to minimize this
contamination.  First, after recovering $\thi$ and $\tovi$ , an
initial correction is made for contamination by higher-order 
\HI\ lines by subtracting 
\begin{equation}
\tau_{{\rm corr}}(\lambda)=\tlya(\lambda^\prime)(f_{{\rm Ly}i}\lambda_{{\rm Ly}i}/f_{\rm Ly\alpha}\lambda_{\rm Ly\alpha}),
\end{equation}
where \textit{f} is the transition oscillator strength, $\lambda
^\prime=(\lambda_{\rm Ly\alpha}/\lambda_{{\rm Ly}i})\lambda$ is the
redshifted Lyman $\alpha$ wavelength corresponding to Ly$i$ absorption
observed at wavelength $\lambda$ and \textit{i} corresponds to first five
higher order Lyman lines (i.e., \lyb\ $\lambda1025$ through
Ly${\zeta}$ $\lambda930$).   

A second correction is made by taking the minimum of the \OVI\ doublet,
\begin{equation}
\tovi={\rm min}\left (\tau_{{\rm OVIa}},\frac{f_{{\rm OVIa}}\lambda_{{\rm
      OVIa}}\tau_{{\rm OVIb}}}{f_{{\rm OVIb}}\lambda_{{\rm
      OVIb}}}\right ) 
\end{equation}
where `a' and `b' denote the stronger and weaker doublet components, respectively.  These corrections are described in detail in Paper I, \S4.2.

Another potential contamination issue is that due to strong \lya\ or
\lyb\ absorption, some higher percentiles in \OVI\ absorption can
become dominated by saturated pixels, so that the particular value of
the percentile is determined by the contaminating lines rather than by
the \OVI\ distribution.  To remove these unreliable percentiles from
consideration,  
the average noise $\bar{\sigma}_{sat}$ is calculated for saturated \OVI\ pixels
and is converted into a maximum optical
depth ($\tau_{{\rm sat}}=-\ln 3\bar{\sigma}_{sat}$).  After \OVI\
optical depths are binned, those percentile bins with
$\tovi>\tau_{{\rm sat}}$ are excluded from the analysis.

\subsection{Testing \OVI\ Recovery}
\label{sec-ovitest}

\begin{figure}
\plotone{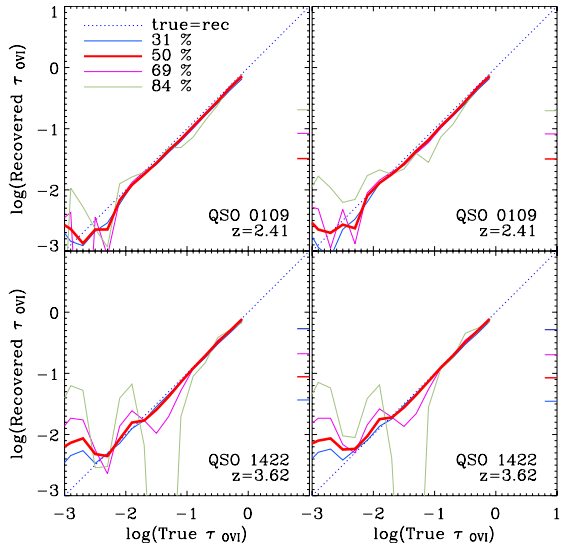} 
\figcaption[]{The accuracy of recovering the ``true'' OVI pixel optical depth from
simulated spectra of Q0109$-$3518 and Q1422$+$230.  For every panel, the
recovered OVI optical depth is plotted against the ``true'' optical
depth, $\ttru$.  Each panel shows the binned percentiles after $\tmin$ 
has been subtracted from the recovered $\tovi$ pertaining to that
percentile.  The left and right panels calculate $\tmin$ using the
$\log \ttru <-3$ and $\log\thi<-1$ respectively.  These are shown as
horizontal dashes on the right axis of each plot.  For $\log \tovi>-2$
the median OVI optical depths are effectively recovered.   
\label{fig:true_rec}}   
\end{figure}

Because there is substantial processing of the recovered \OVI\ optical
depths, it is important to test how efficiently the true \OVI\ optical
depths are recovered by our procedures.  To do so, simulations were
produced just as described in \S~\ref{sec-overview}, but with
(effectively) perfect
resolution, no noise, and only \OVI\ $\lambda1032$ absorption (and
hence no contamination or self-contamination). In Fig.~\ref{fig:true_rec} the
recovered \OVI\ pixel optical depths are plotted against these ``true''
optical depths for a set of 60 simulated spectra, for two
representative QSOs (see Paper I for more such tests).  Of
particular note is the efficacy of subtracting the ``flat level''
$\tau_{\rm min}$ as described in the preceding section. Ideally,
$\tau_{\rm min}$ would  be determined using pixels with negligible \OVI\
absorption; this is possible in the present case (as the true \OVI\
optical depths are known) and corresponding results are shown in the
left panels. In a realistic case, a proxy for \OVI\ must be used; in
the right panels of Fig. \ref{fig:true_rec}, \HI\ is employed and
$\tau_{\rm min}$ is computed using all pixels with $\log\tau_{\rm HI}
< -1$. In both
cases, the subtracted $\tau_{\rm min}$ values are shown as horizontal
dashes on the right axes of Fig.~\ref{fig:true_rec}. 

Overall, we find that using $\thi$ to calculate $\tmin$ is effective
at recovering $\tovi>10^{-2.5}$ for the 31st and median percentiles; for higher percentiles the recovery is accurate only at higher $\tovi$, but the large `scatter' indicates that this is random, rather than systematic error.

\subsection{Ionization corrections}
\label{ioncorr}
In Papers I and II it was shown from simulations that there exists a
tight correlation between $\thi$ and the absorbing gas density and
temperature, which could be used to predict an ionization correction
(i.e., the ratios of \OVI/O and \HI/H) as a function of density; for
details see \S~6 in Paper I and \S~5.1 in Paper II.  As noted in
Papers II and III, this works well for \CIV\ and less well for \SiIV,
due to their mild and strong ionization correction dependence on
$\thi$, respectively.   

\begin{figure}
\plotone{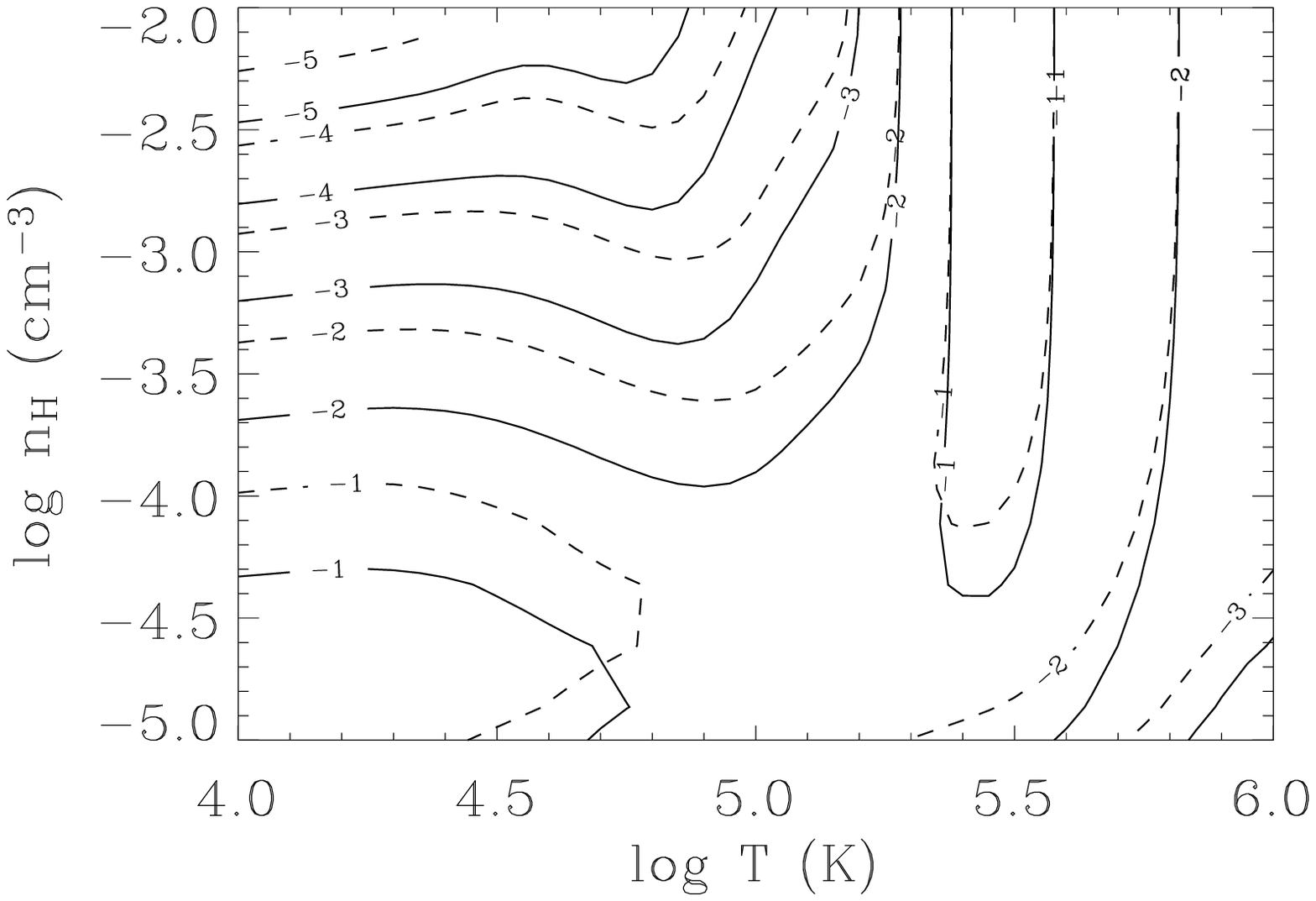} 
\plotone{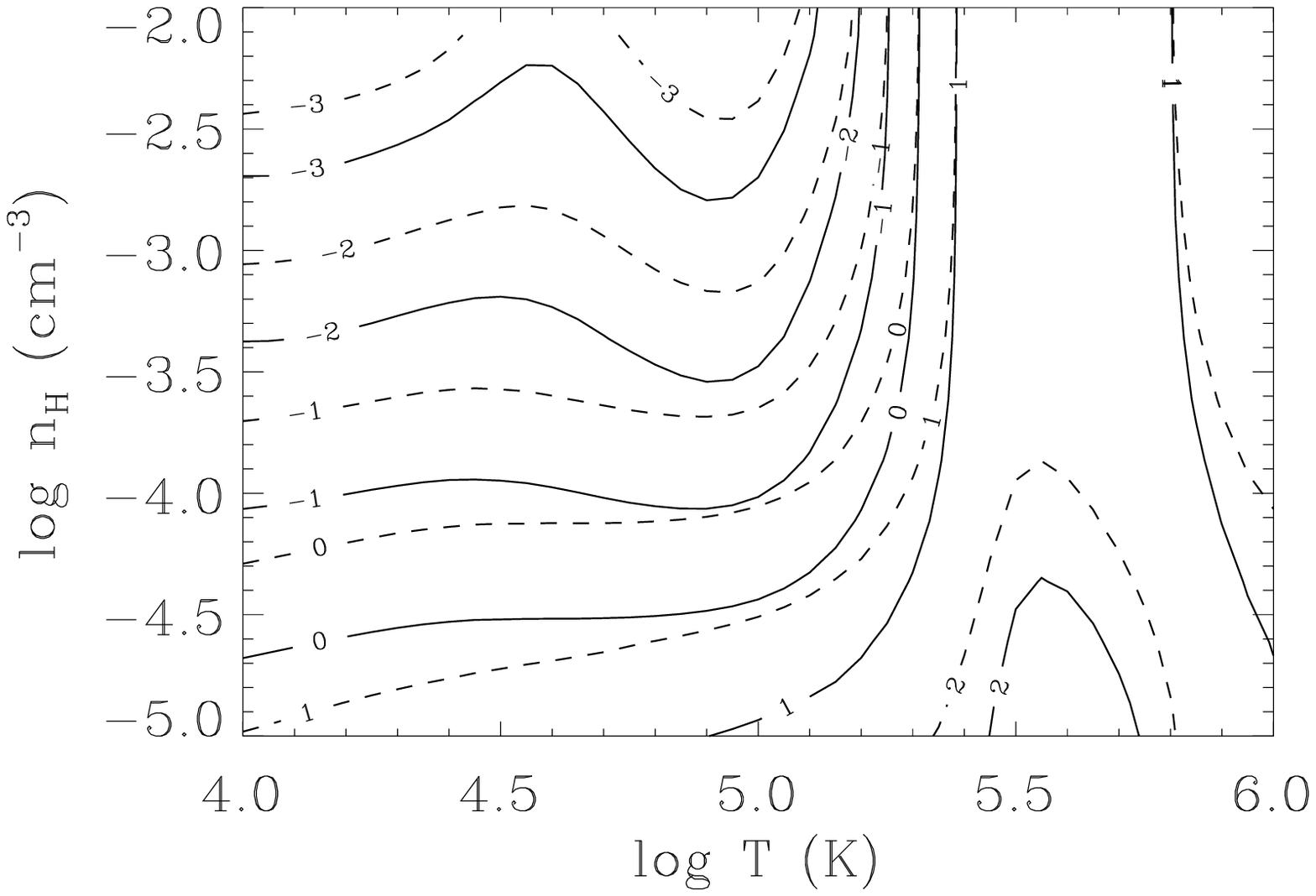} 
\plotone{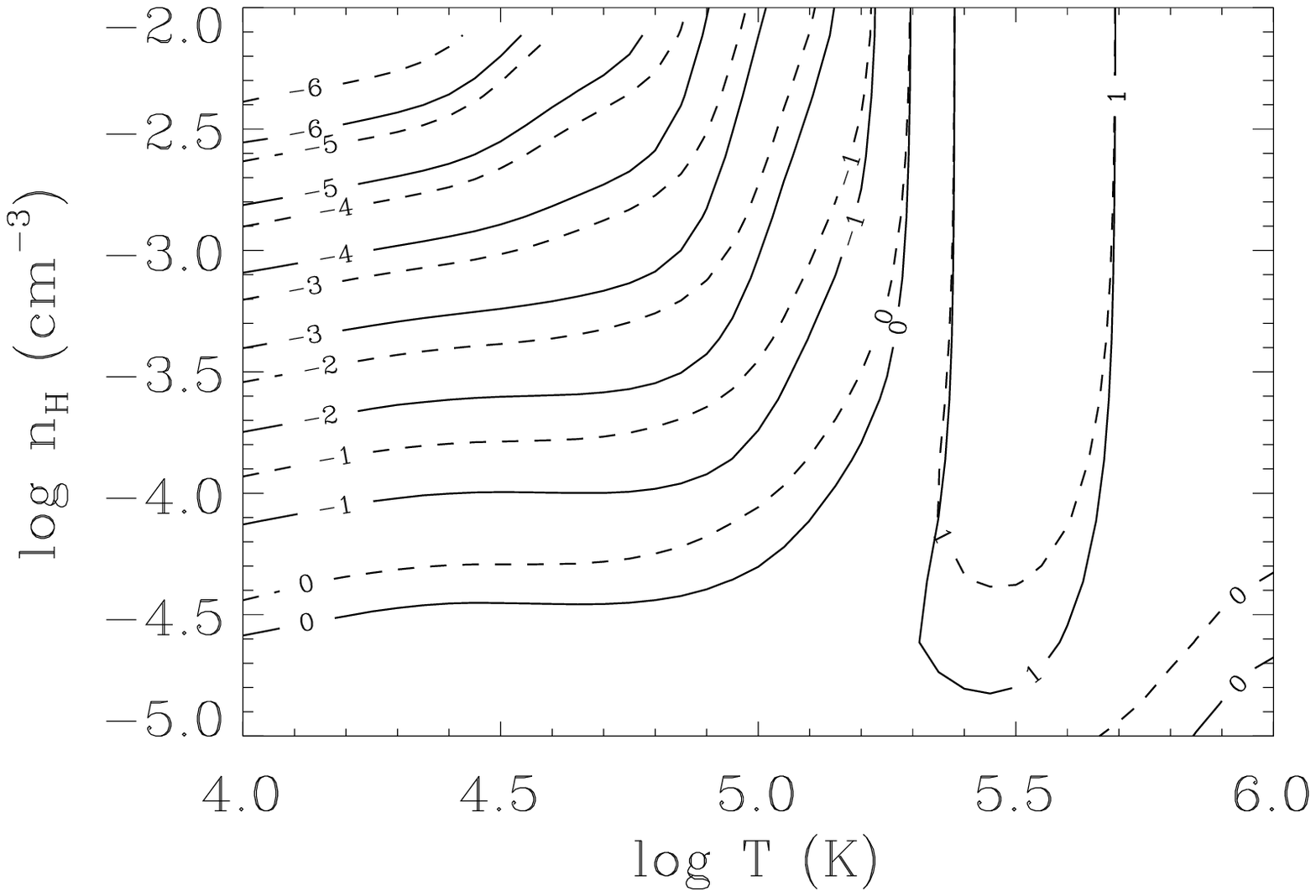} 
\figcaption[]{OVI ion fraction $N_{{\rm OVI}}/N_{{\rm O}}$ (\emph{top}),
  $\log \tovi/\tciv$ for [O/C]=0 (\emph{middle}), and $\log
  \tovi/\thi$ for [O/H]=0 (\emph{bottom}) as functions of temperature
  and the hydrogen 
  number density. Solid (dashed) contours are for the $z=3$ UV background
  model QG (Q).  A measurable quantity of OVI could be
  collisionally 
  ionized if $T\gtrsim 10^{5}~\K$, which would be 
  misinterpreted as photoionized gas during the ionization correction
  process.  This gas would potentially contain very little associated
  CIV and HI. 
\label{fig:ionpred}}
\end{figure}

In the upper panel of Figure \ref{fig:ionpred} we provide a contour plot of
the logarithm 
of the predicted fraction of \OVI\ ions versus temperature and
density.  The middle and lower panels show $\log 
\tovi/\tciv$ for [O/C]=0, and $\log \tovi/\thi$ for [O/H]=0,
respectively.  For photoionized gas ($T \la 10^{5}K$) the \OVI\
fraction is highest for $-5\la \log n_{{\rm H}} \la -4$, and only
weakly dependent on the temperature.   

However, for higher densities (and $T \la 10^5\,$K), the \OVI\
fraction falls quickly, resulting in a very large ionization
correction for \HI\ saturated pixels.  At the same time,  the \OVI\
fraction at high density increases with $T$ for $T \gtrsim 10^5\,$K, so
that collisionally ionized gas might be detected relatively easily.
Therefore at high $\thi$, collisionally ionized \OVI\ gas can easily
swamp photoionized \OVI.  This can be seen most clearly in the
bottom panel, which shows that for fixed [O/H] and density,
$\tovi/\thi$ increases quickly at $T \gtrsim 10^5\,$K, particularly if the
density is high. Because our
$\thi-\rho$ relation is dominated by photoionized gas (with $T <
10^5$), the effects of collisionally ionized gas are potentially
important and are discussed at length in \S~\ref{sec-collisional} below. 

The strategy employed here is to employ our fiducial ionization
corrections as in Papers II and III, but to recognize that at
high-density, the results may be significantly affected by
collisionally ionized gas. It is important to note that the importance
of collisionally ionized gas may be underestimated by our simulation,
because we did not include a mechanism for generating galactic winds,
which could shock-heat the gas surrounding galaxies. Our simulation
does, however, include heating by gravitational accretion shocks.

\begin{figure}
\plotone{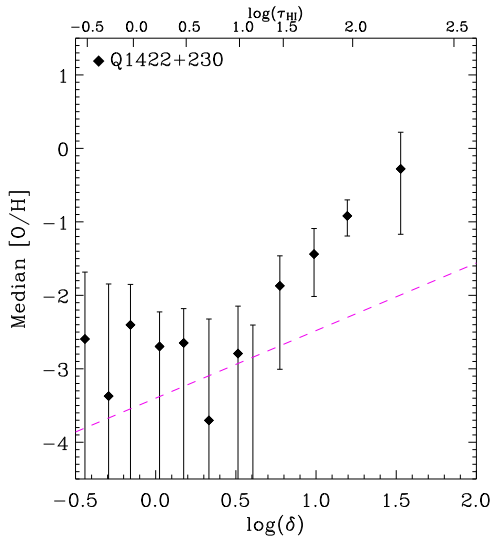} 
\figcaption[]{The median oxygen abundance as a function of the overdensity
  (\textit{bottom axis)} or $\thi$ (\textit{top axis}), by applying
  the ionization correction of \S \ref{ioncorr} and Paper II.  For
  reference, the dashed line is [O/H] vs.\ density, assuming the carbon
  distribution fit for Q1422+230 from Paper II and constant
  [O/C]=0.64. 
\label{fig:1422ioncorr}}
\end{figure}

Once the ionization correction has been determined, and corrected
\OVI\ optical depths and recovered \HI\ optical depths $\thi$ have been
obtained, the oxygen abundance can be calculated: 
\begin{equation}
[{\rm O/H}] 
=  \log \left ({\tau_{\rm OVI} \over \tau_{\rm HI}} 
{(f\lambda)_{\rm HI} \over (f\lambda)_{\rm OVI}}  
{n_{\rm O} \over n_{OVI}}{n_{\rm HI} \over
  n_{H}}\right ) - ({\rm O/H})_\odot,
\label{eq:metallicity} 
\end{equation}
where $f_i$ and $\lambda_i$ are the oscillator strength and rest
wavelength of transition $i$, respectively ($f_{\rm OVI} = 0.1329$,
$f_{\rm HI} = 0.4164$, $\lambda_{\rm OVI} = 1031.9270$~\AA,
$\lambda_{\rm HI} = 1215.6701$~\AA), and we use the solar abundance
$({\rm O/H})_\odot = -3.13$ (number density relative to hydrogen;
Anders \& Grevesse 1989).  An example of the results from this
analysis applied to the observed spectrum of Q1422+230 is shown in
Figure \ref{fig:1422ioncorr}. 

\begin{figure}
\plotone{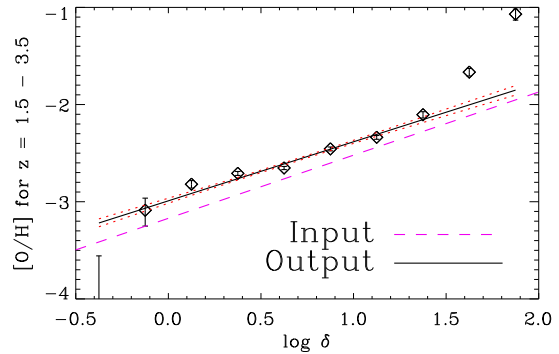} 
\figcaption[]{Test of the OVI ionization correction from the synthetic
  spectra using the full QSO sample (using 50 realizations of each synthetic spectrum so as to focus on systematic effects).  The input line represents the
  OVI distribution imposed on the simulations, and the output line is
  that recovered from the analysis and ionization correction, fit to the data at  $\log \delta\leq$1.5; in this density range, the recovery is accurate to within 0.3 dex. 
\label{fig:trueinvsumm_O6}}
\end{figure}

In Figure \ref{fig:trueinvsumm_O6} we show a test in which we have
generated simulated spectra using the ``QG'' ionizing background,
recovered optical depths, and applied the  
just-described ionization correction to recover the oxygen abundance.
The true metallicity is given by the carbon distribution of Paper II
(for the QG background), with a  
fixed [O/C]=$\log 4.5 \approx 0.65$ (i.e., [O/H]=-3.15+0.65$\delta$),
and is shown 
on the plot as a dashed line.  For $\log \delta \leq 1.5$, the \OVI\
abundance recovery is promising:  it overestimates by less than 0.3
dex, and the dependence of $\delta$ is reproduced.  However, the
overestimation appears to increase for $\log \delta \geq 1.5$,
reaching approximately 1 dex for the highest overdensity bin.  The
difference in the high- versus low-$\delta$ gas can probably be
attributed to the collisionally ionized gas residing in and around
dense regions, due to gravitational accretion shocks. We should thus
keep in mind that we expect to overestimate the oxygen abundance
associated with strong \HI\ absorbers.

\section{Results}
\label{sec-resrel} 

Our basic result will be an estimate of [O/C] for the low-density IGM
as a whole, computed using four different but consistent methods,
which we describe in turn. 

\subsection{$\tovi$ versus $\tciv$ for the full sample}
\label{sec-resrel-oc}

\begin{figure}
\epsscale{1.0}\plotone{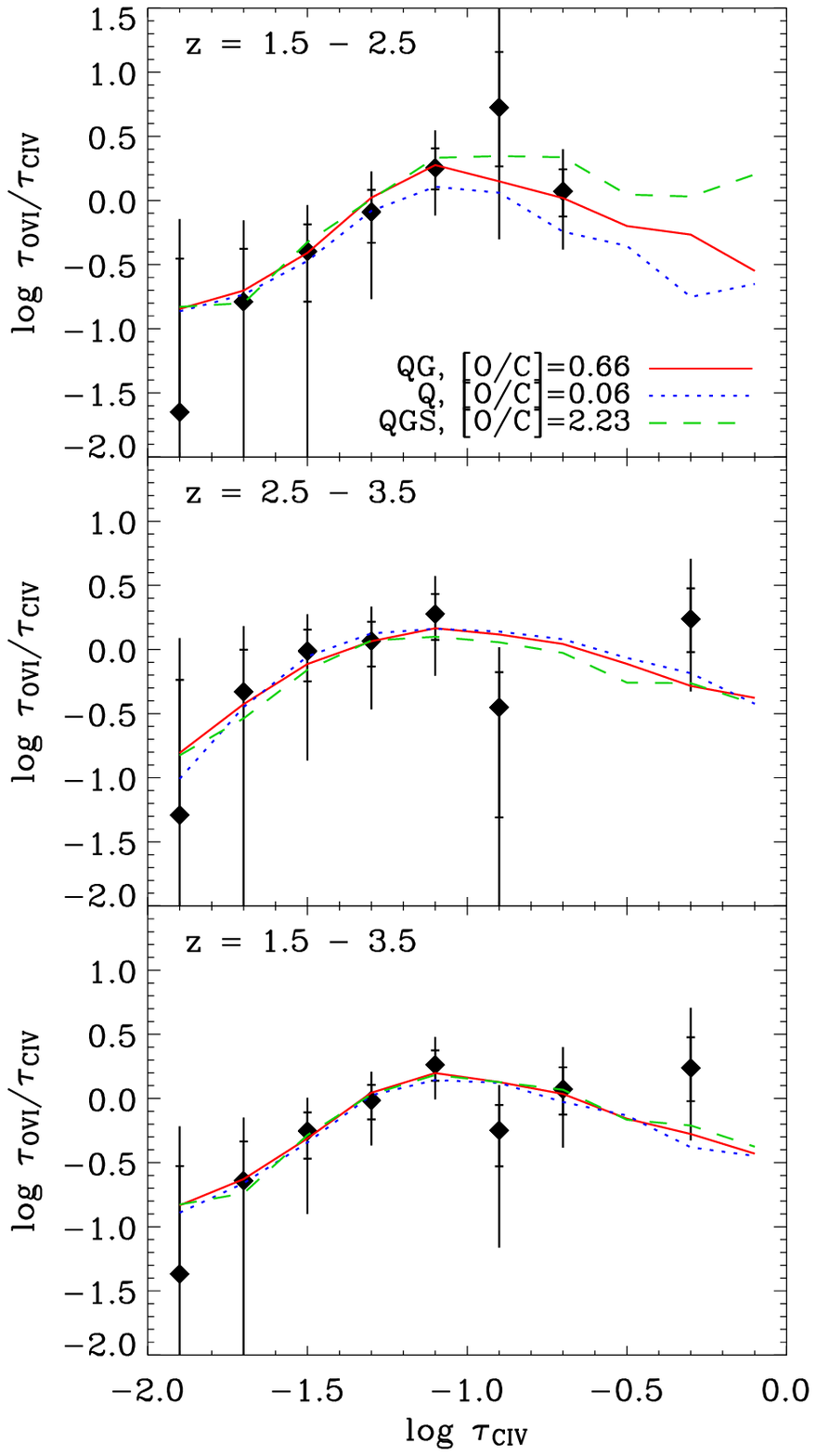} 
\figcaption[]{Rebinned median $\log \tovi/\tciv $ vs.\
$\log\tciv$ in bins of $z$ for the combined QSO sample. The first two
panels show bins centered at $z=2$ and 3 with width $\Delta z=1$;
the bottom panel shows combined data for all redshifts.  Data points
are plotted with 1 and 2$\sigma$ error bars.  The 
lines represent corresponding simulation points (with errors
suppressed and with [O/C] chosen to minimize the $\chi ^2$) using
different UVB models, as indicated in the legend of the top panel.  
\label{fig:O_C.fwd.zbin}}
\end{figure}

To extract as much information as possible from our data we have, as
in papers II and III, combined the data points obtained from our
entire sample.   
Figure~\ref{fig:O_C.fwd.zbin} shows $\log
\tovi/\tciv$ versus\ $\log \tciv$, in bins of $z$. To generate these
points, we begin with $\tovi$ values binned in $\tciv$ for each QSO, as in
Fig.~\ref{fig:oviscat} for Q1422+230.  We then subtract from each the
``flat level'' 
$\tau_{\rm min}$ for that QSO to adjust for noise, contamination, etc.
(see \S~\ref{sec-overview}), then divide by the central value of the
$\tciv$ bin.  These points, gathered from all QSOs, are rebinned by
determining, for each $\tciv$ bin in 
Fig.~\ref{fig:O_C.fwd.zbin}, the best constant-level $\chi^2$-fit to
all of the points in the 
specified redshift bin. The errors represent 1- and 2-$\sigma$
confidence intervals ($\Delta\chi^2=2$ and $\Delta\chi^2=4$) on this fit.

The plotted lines indicate corresponding optical depths from synthetic
spectra drawn from the simulation, using several UVB models, the
corresponding [C/H] distributions as determined in Paper II, and a
constant [O/C] value determined as follows. 
For each background, we generate simulated $\tovi/\tciv$ points in the
same way as we did for the observations, but averaging over 50
simulated realizations as described in~\S~\ref{sec-overview}.  We then
calculate a $\chi^2$ between all valid observed original (not
rebinned) points and the corresponding simulated points.\footnote{Even
using 50 realizations, it may occasionally happen that a simulated
bin fails to have enough pixels for at least five realizations, and
so is undefined; in this case the observed point is discarded as
well.}  Because we use 50 simulated realizations, the simulation
errors are almost always negligible compared to the observed errors,
but they are still taken into account by calculating the total $\chi^2$
using the formula:
\begin{equation}
\chi^2=\sum_i \left[\left({X_{\rm obs}-X_{\rm sim}\over\sigma_{\rm
      obs}}\right)^{-2}+\left({X_{\rm obs}-X_{\rm sim}\over\sigma_{\rm
      sim}}\right)^{-2}\right]^{-1},
\label{eq:chi2}
\end{equation}
where $X\equiv\tovi/\tciv$ and $\sigma$ is the error in this
quantity.  We then add a constant offset to the simulated points
(which corresponds to scaling [O/C]) such that $\chi^2$ is
minimized.  In each panel the lines connect the scaled, rebinned
simulation points.

\begin{figure}
\epsscale{1.0}\plotone{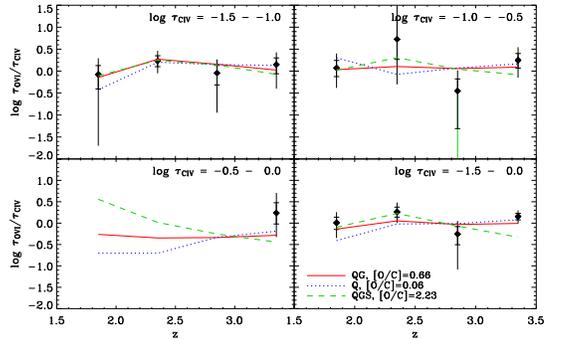} 
\figcaption[]{Rebinned median $\log(\tovi/\tciv)$ vs.\ $z$ in cuts of $\tciv$
for the combined QSO sample. Data points
are plotted with 1 and 2$\sigma$ error bars, where green lines denote
lower error of -$\infty$. The first three panels show bins centered
at $\log\tciv=-1.25,-0.75$ and -0.25 with width $0.5$, dex; the
bottom-right panel shows data for all $\tciv$ combined.  The lines are
the same as in Fig.~\ref{fig:O_C.fwd.zbin}.
\label{fig:O_C.fwd.dbin}}
\end{figure}

The first evident result is that 
$\log\tovi/\tciv\sim 0$ and appears to be at most weakly dependent on 
$\log\tciv$, from $\log\tciv~-1.5$ to
0.  This is unlike $\tsiiv$, which increases by $\approx$2 dex in this
$\tciv$ regime (Paper II).  At the lowest densities the data exhibits
a decline.   
Comparing these panels suggest there is little dependence on redshift
in this interval.  This can be seen more clearly in
Figure~\ref{fig:O_C.fwd.dbin}, which show $\log \tovi/\tciv$ versus
$z$ in bins of $\tciv$: there is no evidence, in either the simulated
or observed points, for evolution in $\tovi/\tciv$ for 1.5$\leq z
\leq$3.5.

The observed trends in $\log(\tovi/\tciv)$ are reproduced well by the
simulations. Therefore, because $\tovi/\tciv$ scales with [O/C], the offset in
$\tovi/\tciv$ obtained by minimizing the 
$\chi^2$ (Eq.~\ref{eq:chi2}) against the observations can be used to
reliably compute the best fit [O/C].  As an example, for our fiducial UVB model QG, the
simulated spectra were generated with [O/C]=0.65 and best fit by an offset 
of $+0.01$\,dex (implying a best-fit [O/C]=0.66), with $\chi^2/{\rm
  d.o.f.}$=42.7/82.  As we found in Paper II and for Q1422+230 above,
the reduced $\chi^2$ is somewhat low.
This is due to a slight overestimate of the errors at low-$\tciv$
(Paper II) and to the fact that the data points are not completely
independent (because single absorbers contribute to multiple data
points). 

The fitted [O/C] values and corresponding $\chi^2/{\rm d.o.f.}$, are
listed in Table~\ref{tbl:allfits}, with errors
computed by bootstrap-resampling the quasars used in the $\chi^2$
minimization.  For our fiducial model, QG, the best fit
[O/C]$=0.66^{+0.06}_{-0.06}$.  The quasar-only background Q (which is
probably too hard; see Paper II) gives a much lower value of
[O/C]$=0.06^{+0.06}_{-0.06}$.  The softer QGS
backgrounds gives implausibly high values of
[O/C]$=2.23^{+0.06}_{-0.06}$. Coupling this with results from Paper II
suggesting that the QGS background is unrealistically soft, strongly
disfavors this UVB model. 

\begin{deluxetable*}{lcccccccccc} 
\tabletypesize{\scriptsize}
\tablewidth{0pc} \tablecaption{Best-fit Abundances
\label{tbl:allfits}}  
\tablecomments{The best-fit abundances and corresponding $\chi^2$/dof
  from this work, and Papers II ([C/H]) and III ([Si/C]).  Errors were
  computed by bootstrap-resampling the quasars used in the $\chi^2$
  minimization.  } 
\tablehead{ & 
&&&&&&
\multicolumn{4}{c}{[C/H]=$\alpha$+$\beta(z-3)$+$\gamma(\log\delta - 0.5)$} \\
\colhead{UVB model} & 
\colhead{[O/C]\tablenotemark{a}} & \colhead{$\chi^2$/d.o.f.} &
\colhead{[O/C]\tablenotemark{b}} &
\colhead{[Si/C]} & \colhead{$\chi^2$/d.o.f.} &
\colhead{[Si/O]} &
\colhead{$\alpha$} & \colhead{$\beta$} & \colhead{$\gamma$}  & \colhead{$\chi^2$/d.o.f.}}
\startdata
QG & $0.66 \pm 0.06$ &  42.7/82 & $0.56\pm 0.08$ & $0.77 \pm 0.05$ & 65.7/115 & $0.11 \pm 0.08$ & -3.47$^{+0.07}_{-0.06}$ & 0.08$^{+0.09}_{-0.06}$ & 0.65$^{+0.10}_{-0.14}$ & 114.1/184\\
Q & $0.06 \pm 0.06$ &  59.9/82 & $0.06 \pm 0.08$ & $1.48 ^{+0.05}_{-0.06}$ & 65.6/115 & $1.42 \pm 0.08$ & -2.91$\pm 0.07$ & -0.06$\pm 0.09$ & 0.17$\pm 0.08$ & 113.8/184\\
QGS & $2.23\pm 0.06$ & 26.7/82 & $2.11\pm 0.09$ & $0.26 ^{+0.06}_{-0.07}$ & 73.8/115 & $-1.97 \pm 0.11$ & -4.14$^{+0.06}_{-0.05}$ & 0.54$^{+0.10}_{-0.07}$ & 1.31$\pm 0.07$ & 114.2/184\\
\enddata
\tablenotetext{a}{Computed using $\tovi/\tciv$ vs.\ $\tciv$; see
  \S~\ref{sec-resrel-oc}.} 
\tablenotetext{b}{Computed using $\tovi/\thi$ vs.\ $\thi$ for gas of density $\delta \le 10$; see \S~\ref{sec-ovihi}.} 
\end{deluxetable*}

We may also subdivide our sample by redshift to test the
dependence of [O/C] on these.  First, computing [O/C] using only
spectra that have a median absorption redshift ${\rm med}(z) > 2.5$
(see Table \ref{tbl:sample}) yields [O/C]=$0.71 \pm 0.07$, versus
[O/C]=$0.58 \pm 0.10$ using the spectra with ${\rm med}(z) <
2.5$; these are consistent to about 1$\sigma$.  Using the Q and QGS UVBs, the [O/C] values inferred from the redshift subsamples are marginally consistent with each other and with the full sample.  

\subsection{$\tovi$ versus  $\thi$ for the full sample}
\label{sec-ovihi}

While the $\tovi/\tciv$ ratios give the most direct constraints on
[O/C], it is also useful to examine $\tovi/\thi$, since comparing the
simulated to the observed $\tovi/\thi$ ratios gives an additional (but
related) estimate of [O/C] (recall that our simulation reproduces the
observed $\tciv(\thi)$)..   

\begin{figure}
\epsscale{1.0}\plotone{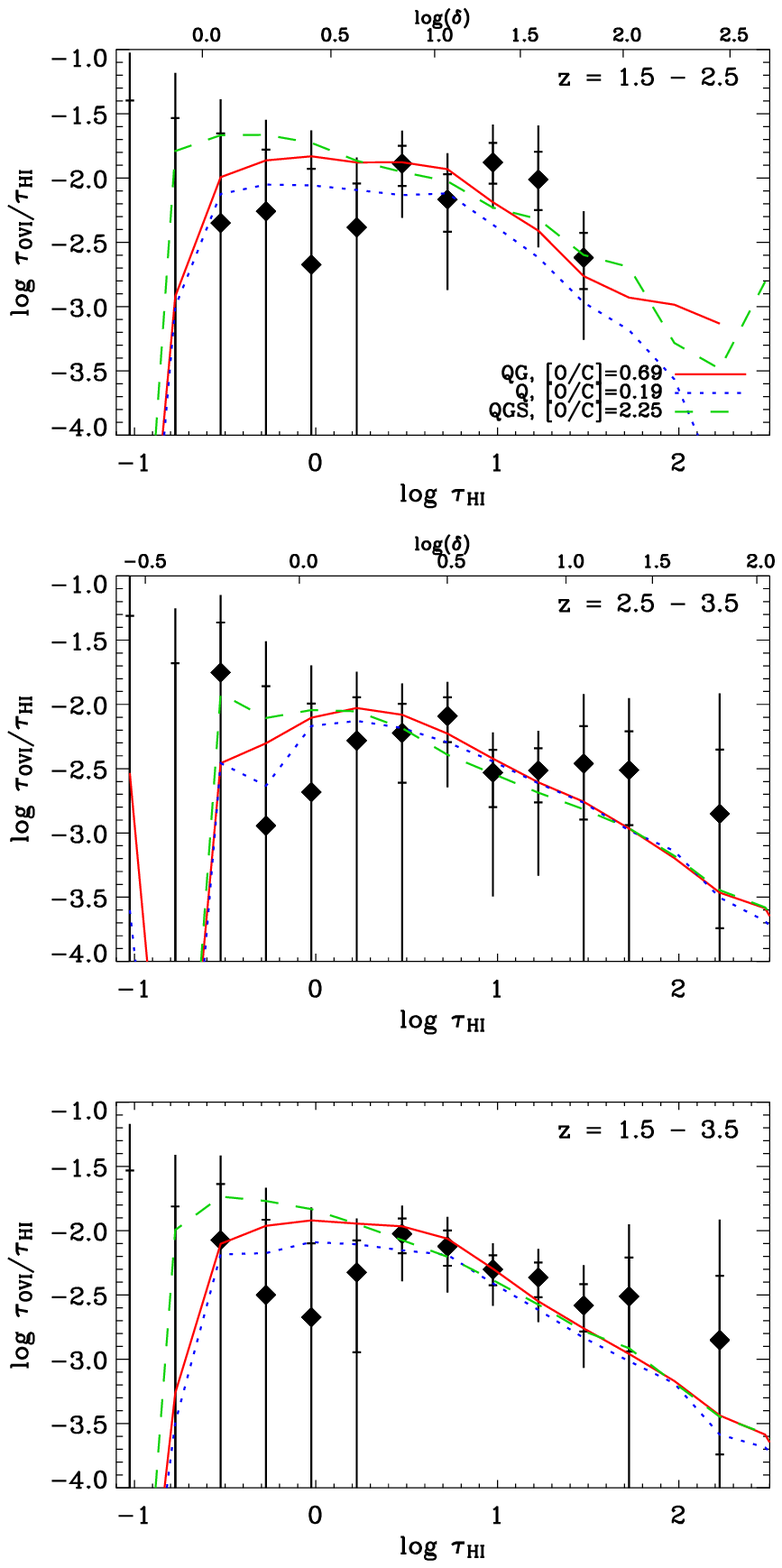} 
\figcaption[]{Rebinned median $\log \tovi/\thi $ vs.\
$\log\thi$ in bins of $z$ for the combined QSO sample. The top two
panels show bins centered at $z=2$ and 3 with width $\Delta z=1$;
the bottom panel shows combined data for all redshifts.  Data points
are plotted with $1\sigma$ and 2$\sigma$ error bars.  The 
lines represent corresponding simulation points (with errors
suppressed and with [O/C] chosen to minimize the $\chi^2$) using
different UVB models, as indicated in the legend in the top panel.  
\label{O_H.fwd.zbin}}
\end{figure}

Figure~\ref{O_H.fwd.zbin} shows $\log \tovi/\thi$ versus $\thi$ for our combined sample, in bins of redshift.  Lines again connect the simulation points (with an overall scaling to best match the observations), which reproduce the observed trends in $z$ and $\thi$.  The scalings correspond to best-fit [O/C] of 0.69 $\pm$ 0.06, 0.19 $\pm$ 0.06, and 2.25 $\pm$ 0.06, for QG, Q, and QGS, respectively.  For QG and QGS, the inferred [O/C] are consistent with the results found using $\tovi/\tciv$; for our hardest UVB model, Q, the [O/C] values are more discrepant, but still within 1.5$\sigma$.

\begin{figure}
\epsscale{1.0}\plotone{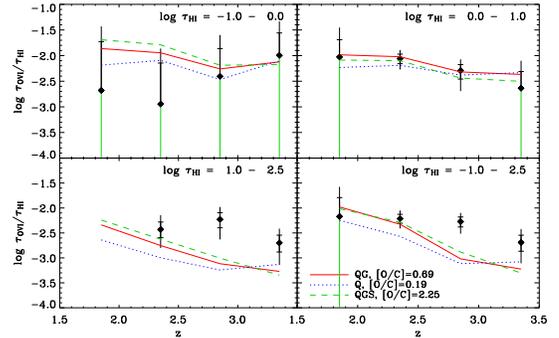} 
\figcaption[]{Rebinned median $\log(\tovi/\thi)$ vs.\ $z$ in cuts of $\thi$
for the combined QSO sample. Data points
are plotted with 1 and 2$\sigma$ error bars, where green lines denote
lower error of -$\infty$. The top two panels show bins centered
at $\log\thi=-0.50$ and 0.50 with width $1.0$~dex; the bottom-left
panel corresponds to a bin centered at $\log\tciv=1.75$ with width $1.50$ dex;
the 
bottom-right panel shows data for all $\tciv$ combined.  The lines are
the same as in Fig. \ref{O_H.fwd.zbin}  
\label{O_H.fwd.dbin}}
\end{figure}

However, while the simulations reproduce the overall trends present in the
data, there are some possible discrepancies. Although the $\log
\thi\leq 0.3$ points are upper limits, the simulations also appear to
fall marginally above the data in this regime. At $\log\thi\gtrsim 1$, on
the other hand, the simulations slightly but significantly
underpredict $\tovi$. This can be seen more clearly in
Figure~\ref{O_H.fwd.dbin},  
where  $\log \tovi/\thi$ versus $z$ in bins of $\thi$ is shown,
exhibiting a clear discrepancy for points at high-$z$ and high
$\thi$. 
Indeed, if we consider subsamples above and below $\log\thi =
1$, we find that for $\log \thi<$1 we obtain [O/C] values consistent
those obtained from $\tovi/\tciv$ (0.62 $\pm$ 0.07, 0.13 $\pm$ 0.07
and 2.18 $\pm$ 0.07 for QG, Q, and QGS respectively), but for $\log
\thi>1$ we obtain [O/C] of $0.92 \pm 0.10$, $0.34 \pm 0.10$, and $2.52
\pm 0.11$ for QG, Q, and QGS respectively. This  discrepancy in [O/C]
for between low- and high-$\thi$ subsamples is significant at $\approx 2.5\,\sigma$ for the QG and QGS models, and at $1.7\,\sigma$ for the Q model.

Given the tight (but $z$-dependent) relation between $\thi$ and gas
density (see Paper II, Fig. 2, and the upper axis in the top two
panels of Fig.~\ref{O_H.fwd.zbin}), it is useful also to divide our
sample into high- and low-density subsamples.  We have done this by
recomputing [O/C] from $\tovi/\thi$ using only bins with $\thi$
corresponding to $\delta < 10$ or $\delta > 10$.  For $\delta<$10, we
obtain [O/C] of 0.56 $\pm$ 0.08, 0.06 $\pm$ 0.08, and 2.11 $\pm$ 0.09
(all consistent at 1$\sigma$ with the full sample), whereas for  for
$\delta>10$ we obtain [O/C] of 1.02 $\pm$ 0.09, 0.51 $\pm$ 0.09, and
2.60 $\pm$ 0.09  for QG, Q, and QGS, respectively (all discrepant at
$\approx 4\sigma$).   

This high-$\delta$ and (less significantly) high-$\thi$ difference
might have several causes.  First, it might correspond to a genuine
change in [O/C] with gas density. In this case, however, such an
effect would also be expected in the trends of $\tovi/\tciv$ vs.\
$\tciv$ -- and no such effect is evident. 
Thus, we
consider it more likely that there exists a significant portion of hot
gas -- not present in the simulations -- that contains \OVI,
but lacks \HI\ and \CIV.  
As suggested in \S\ref{ioncorr} and
Fig.~\ref{fig:ionpred}, and further discussed below in  \S
\ref{sec-collisional}, this might indicate the presence of a
significant amount of collisionally ionized ($T>10^5\,$K) gas in the
IGM.  

\subsection{$\tovi$ versus  $\tsiiv$ for the full sample}
\label{sec-ovisiiv}

\begin{figure}
\plotone{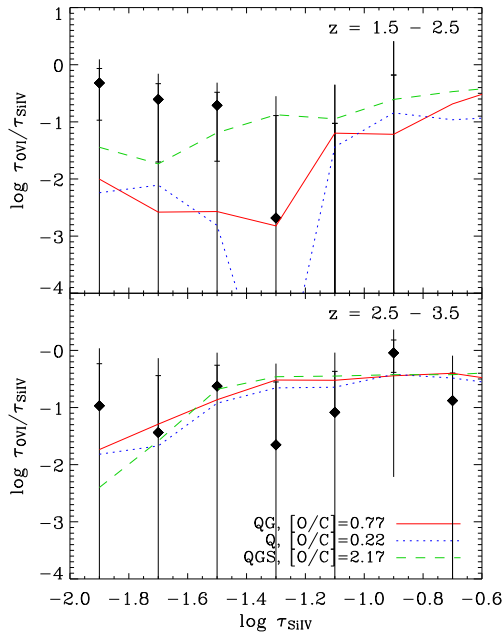} 
\figcaption[]{Rebinned median $\log \tovi/\tsiiv $ vs.\
$\log\tsiiv$ in bins of $z$ for the combined QSO sample. The two
panels show bins centered at $z=2$ and 3 with width $\Delta z=1$.
Data points are plotted with 1 and 2$\sigma$ error bars.  The 
lines represent corresponding simulation points (with errors
suppressed and with [O/C] chosen to minimize the $\chi^2$) using
different UVB models.  The $\log \tsiiv =-1.35$ Q data point is
set to the minimum allowed $\tovi$, $\log \tovi$=-6.
\label{O_S.fwd.zbin}} 
\end{figure}

A third check on our measured [O/C] is provided by the $\tovi/\tsiiv$
ratio.  Fig.~\ref{O_S.fwd.zbin} shows $\tovi/\tsiiv$ versus $\tsiiv$
for our full sample, for two cuts in $z$. For each UVB the constant
Si/C ratio derived in Paper III and the C distribution measured in
Paper II are imposed on the simulations and the [O/C]
ratio is varied to minimize the $\chi^2$ difference between the
observed and simulated data points. Though the
detection of $\tovi/\tsiiv$ is weak, the simulations appear to
adequately represent the observations in the redshift range where
$\tsiiv$ is best detected, $2.5\leq z \leq 3.5$.  From these fits we
infer, for this redshift interval, [O/C] of  $0.77 \pm 0.19$, $0.24
\pm 0.20$, and $2.17 \pm 0.21$ 
for UVB models QG, Q, and QGS, respectively; all are consistent with  the results obtained from
$\tovi/\tciv$, though all are somewhat higher. Because \SiIV\ probes higher density gas
than \CIV\ (see Paper III), this again suggests that the simulations
underpredict the amount of \OVI\ in and near dense gas.

\subsection{[O/H] versus $\delta$ from $\tovi/\thi$ vs.\ $\thi$}
\label{sec-ioncorr}
As a final method, we can apply the ``inversion method'' developed in
Paper II, to convert $\tovi/\thi$ vs.\ $\thi$ into [O/H] vs.\ $\delta$
by applying a density-dependent ionization correction. Then, using the
measured distribution of carbon from Paper II, an independent
measurement of [O/C] can be obtained.

\begin{figure}
\plotone{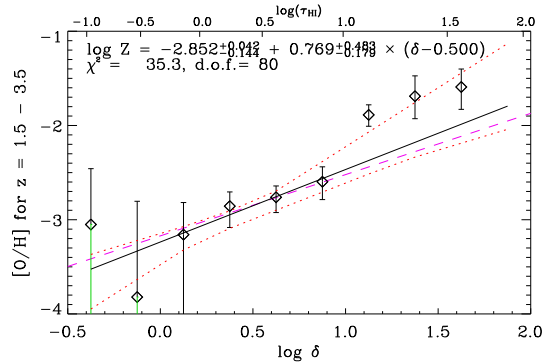} 
\figcaption[]{Median oxygen abundance vs.\ overdensity $\delta$ for
  $z$=1.5-3.5.  The inversion results for each QSO are binned in
  density bins of 0.125 dex.  Data points are plotted with 1$\sigma$
  error bars, where green lines denote lower error of -$\infty$. The solid
  line indicates the least-squares fit to the individual data points for
  $\delta < 10$.  Dotted curves represent the 1$\sigma$ confidence
  limits, which were computed by bootstrap resampling the QSOs. The
  dashed line is the measured [O/H] utilizing the derived [O/C]
  results obtained from $\tovi/\tciv$ and assuming the [C/H]
  distribution from Paper II.  
\label{O_H.inv.zbin}}
\end{figure}

In Figure~\ref{O_H.inv.zbin} we show the derived [O/H] versus $\delta$
for our preferred UVB model QG, with data from all $z$ combined.  The
data points from the individual quasar spectra (an example is shown in
Fig.~\ref{fig:1422ioncorr}) have been binned in density
bins of 0.25 dex.  The solid line shows the least-squares fit to the
data points with $\delta < 10$ and the dotted curves indicate the 1
$\sigma$ confidence 
limits, with the resulting fit given in the upper left corner.  The
errors on the fits were determined by bootstrap resampling the QSOs.
The dashed line is the value of [O/H] given by the derived
[O/C]=$0.64$ result using $\tovi/\tciv$, and assuming a [C/H]
distribution from Paper II.   

For $\delta< 10$, the [O/H] derived from the ionization correction
agrees very well with that determined from both $\tovi/\tciv$, and
$\tovi/\thi$.  This strengthens the result from \S\ref{sec-resrel-oc}
and \S\ref{sec-ovihi} that [O/C] is indeed constant for -0.5$\leq \log
\delta \leq$ 1. 

However, for $\delta> 10$ the ionization correction results in
substantially more [O/H] than that predicted using the
previously-determined [C/H] distribution and a fixed [O/C] ratio.
This is similar to the breakdown between the simulations and observed
$\tovi/\thi$ seen in Fig.~\ref{O_H.fwd.zbin}, and the erroneously high
[O/H] recovered for high-$\delta$ (see Fig.~\ref{fig:trueinvsumm_O6})
when applying the ionization correction to the simulated spectra.
Once again, this suggests the presence of collisionally ionized \OVI. 

\subsection{Systematic uncertainties}

The primary source of systematic uncertainty in, e.g. [O/C] or [O/H], is the complex modeling that must be performed to extract these values from the pixel correlations.  
The greatest combination of importance and uncertainty is clearly the uncertainty in the shape of the UVB; but this is discussed at length below in Sec.~\ref{sec-nuc} so we here focus on other aspects of the modeling.

The good agreement between the four methods we have employed indicates that the method is sound; but it is also clear that there are real differences between the universe and our simulation.  In particular, it is clear that the real universe has an extra component of \OVI\ at high density, almost certainly due to collisionally ionized gas that is not captured by those simulations.  Nonetheless, if we exclude those high-$\thi$ regions, the small discrepancies between [O/C] as measured using the different methods indicate that such effects probably do not contribute more than $\sim 0.1\,$ dex uncertainty to our basic results.
	
Another source of error that may be inaccurately assessed by our
bootstrap resampling technique is that from continuum fitting.  As
shown in Table~\ref{tbl:sample}, our estimated rms error in the
\OVI\ absorption region is $\sim 1-4\,$\%.  To test the effect of this
error on our results, we have imposed an additional error on each
observed spectrum on scales of 20, 80, and 320\,\AA, for a total added
rms error of 2\%, then recomputed our results.  We find that our best
fit [O/C] from $\tovi$ versus $\tciv$, and from $\tovi$ versus $\thi$,
are both within $\sim 0.03$~dex. Further, both the individual binned
points and linear fits of the [O/H] values computed from the full
sample (Fig.~\ref{O_H.inv.zbin}) are all affected by this continuum
error to a lesser degree than the quoted random errors.  Thus, we
conclude that the continuum fitting error is not a significant systematic affecting our results.

A final possible source of systematic error is that the \OVI\ and \CIV\ recombination rates used in the version of CLOUDY we have employed are too high by $\sim 50-75$\% compared with recent experimental values, for the temperatures relevant to low-density photoionized gas. (Savin, {\em private communication}).  For the range of densities $1 \lesssim \delta \lesssim 40$ we cover, this would imply a (density-dependent) correction of  $0-0.1$\,dex.  This might change our overall fits by an amount comparable to the statistical errors, but is still much smaller than uncertainties stemming from the UVB shape, and cannot account for the ~0.5\,dex of excess in $\tovi$ at high $\thi \gtrsim 10$ shown in Fig.~\ref{O_H.inv.zbin}.
Combining these possible sources, we estimate probable systematic
errors of $\sim 0.2\,$dex in our basic [O/C] and [O/H] values; this
uncertainty may be somewhat greater in subsets of the data, particularly at high-density.

\subsection{Densities and volume/mass fractions to which the results apply}

It is important to emphasize that each quoted result is sensitive to, and applies to, only a certain range of gas densities.  At the upper end, our results nominally concern gas of up to $\delta \sim 100$, though (as noted at length above) the high-$\thi$ range of our data is likely to be affected by collisional ionization.  That range is, however, not dominant: we have checked that if all pixels with $\thi > 30$ are excluded from the analysis, the results given for [O/C] in columns 2 and 4 of Table~\ref{tbl:allfits} change only within the quoted errors.

The lower end of the gas density range probed is most straightforward in results from $\tovi/\thi$ (Figs.~\ref{O_H.fwd.zbin} and ~\ref{O_H.inv.zbin}, which formally show OVI detections at 1$\sigma$ using all QSOs at z=$2.5-3.5$ for $\log\thi \gtrsim   -0.1$, or $\log\delta \gtrsim  0.1$ and confident detections at $\log\delta \gtrsim  0.4$.  In Paper III, the Si abundance results were sensitive to 
$\log\thi \gtrsim  0.2$ $(\log\delta \gtrsim  0.2)$, and in Paper II, C abundances were measured in 
much lower density gas.  Thus our results provide indirect constraints on [O/C] and [O/Si] down to $\log\delta \sim 0.1-0.2$.  

However, there are important caveats. First, the quoted results pertain to the the full density range probed and thus are not necessarily very sensitive to the lowest densities.  Second, because the pixel method only works if the element on the x-axis is more easily detectable than the element on the y-axis, direct measurements of [O/C] and [O/Si] from $\tovi/\tciv$ and $\tovi/\tsiiv$ are dominated by much higher-density that gives $\tciv,\tsiiv \gg -2.0$; thus while these results are consistent with our indirect constraints, they do not address the (somewhat implausible) possibility that at low densities O, Si, and C come from completely different gas phases.  On the other hand, at {\em high} densities it is quite possible that OVI and SiIV emission are dominated by different phases, so the indirectly inferred [Si/O] values are probably both reliable and well-measured only in the moderate density range $\delta \sim 5-10$.

Although mass or volume filling-factors corresponding to these results are not well constrained (see, e.g., Schaye \& Aguirre 2005), the forewarned reader can convert the density range $\delta \gtrsim  2$ correspond to into a volume using Paper II.

\section{Analysis and discussion of results}
\label{sec-discuss}

\subsection{Relative abundances and the spectral shape of the UVB}
\label{sec-nuc}

The best fitting metallicities and corresponding $\chi^2$/d.o.f.\ from
Papers II, III and this work are shown in Table~\ref{tbl:allfits} for
each UVB model.  Two interesting results stand out. First, for all UVB
models carbon is underabundant relative to both silicon and oxygen (being only marginally
consistent with solar for UVB model Q). Second, all abundance ratios are
sensitive to the UVB shape. A harder UVB results in a lower inferred
[O/C] but a higher inferred [Si/C], making the [Si/O] ratio
particularly sensitive to the spectral hardness of the UVB. 

The extreme sensitivity of the inferred [Si/O] ratio to the
spectral shape of the UV background makes it
possible to constrain feasible UVB models by making only weak
assumptions about the [Si/O] ratio. Since Si and
O are both $\alpha$ elements, they are expected to trace each other
relatively well. For example, using the nucleosynthetic yields of
\cite{1998A&A...334..505P} and \cite{2001A&A...370..194M} and a
\cite{2003PASP..115..763C} initial mass function from $0.1-100~{\rm M}_\odot$, the [Si/O] ratio of 
the ejecta of a population of age $t \gtrsim 10^8$~yr is predicted to be
about 0.12 and -0.03 for stars of solar and 1 percent solar metallicity,
respectively. This agrees well with observations of metal-poor stars,
which find [Si/O]$\approx 0$ \citep{Cayrel2004}. 

Tallying the results
of this work with those of Papers II and III yields [Si/O]$= 0.11\pm
0.08$, $1.42\pm 0.08$, and $-1.97\pm 0.1$ for UVB models QG, Q, and
QGS, respectively. Thus, our preferred model, QG, is nicely consistent
with the expectations, but models Q and QG lead to inferred [Si/O]
ratios that are highly inconsistent with both nucleosynthetic yields
and observations of metal-poor stars. (Assuming [Si/C] $\simeq 0.5$ and [Si/O]  $\simeq 0$ in the Q background, for example, raises the $\chi^2$ of the fits in Figures~\ref{fig:O_C.fwd.zbin} and~\ref{O_H.fwd.zbin} by 65 and 92, respectively; requiring this for QGS likewise raises the $\chi^2$ by 138 and 81.)
We conclude that the UVB has a
spectral shape similar to that of model QG.

While our result using the QG UVB are broadly consistent with the
abundance ratios in metal-poor stars and in yield calculations, the
[O/C] and [Si/C] may be somewhat high, by $\sim 0.1-0.3$\,dex. This is
comparable to our systematic errors, but nevertheless interesting if
taken seriously.   

For example, the models of \citet{Nomoto2006} that include the
contributions of hypernovae (defined as supernovae with kinetic energy
$> 10 \times$ that of normal core-collapse SNe) produce [O/C]
$\approx$ 0.6, and [Si/C] $\approx$ 0.65, in agreement with our
results. 

\subsection{Implications for cosmic abundances}
\label{sec-abun}

In Paper II we combined the median [C/H]$(\delta, z)$ with the width
$\sigma([{\rm C/H}])(\delta,z)$ of the lognormal probability
distribution of [C/H] for $-0.5 \le \log\delta \le 2.0$ to determine
the mean C abundance versus $\delta$.  This was then integrated over the
mass-weighted probability distribution $\delta$ (obtained from our
hydrodynamical simulation) to compute the contribution by gas in this
density range to the overall mean cosmic [C/H].  Assuming that [O/C]
is {\em constant} over this density range\footnote{Note that this is an
extrapolation beyond the range $2\lesssim \delta \lesssim 10$ over which we have reliably {\em measured} Oxygen abundances.} we obtain, for our fiducial UVB model QG, [O/H]$= -2.14 \pm
0.14$, corresponding to
\begin{equation}
\Omega_{\rm O,IGM} \simeq 3.3\times10^{-6} 10^{[{\rm O}/{\rm H}]+2.1}\left ({\Omega_b
\over 0.045}\right ).
\end{equation}
 Extrapolating our [C/H] and [O/C]
results even further to the full density range of the simulation would yield values
$\approx 0.2$\,dex higher but with more uncertainty, as we have argued
that our results are unreliable at the highest densities. 

Note that these results are relatively insensitive to the UVB (unlike those for
$\Omega_{\rm Si, IGM}$ in Paper III) because for  a harder UVB, the
inferred [C/H] increases, while [O/C] decreases.  For our quasar-only
model Q, these effects almost entirely cancel, yielding [O/H]$\approx
-2.3$, and an $\Omega_{\rm O,IGM}$ value 30$\%$ lower than for model
QG.    
Note also that
these estimates include the oxygen that resides in gas that is
observable in \CIV\ and \OVI, but they do not include oxygen in
intergalactic gas that is very hot ($T\gg 10^5$~K) or very cold ($T
\ll 10^4$~K) and shielded from ionizing radiation.

If, following ~\citet{bouche}, we take $\Omega_{\rm Z,
  IGM}=\Omega_{\rm O, IGM}/0.6$, we then infer an intergalactic metal
reservoir of 
$
\Omega_{\rm Z, IGM} \approx (4.3-6.8)\times 10^{-6}.
$
This can be compared to their estimate of the total $z \approx 2-3$
  ``metal budget" of $\Omega_Z \sim 2-3\times 10^{-5}$, indicating
  that $\sim 15 - 35\,\%$ of metals produced prior to $z =3$ reside in
  the component of the IGM that is studied here.

\subsection{Previous Searches for Oxygen in the IGM}
\label{sec-compare}

Previous studies have explored oxygen abundances in the IGM using both
line-fitting~\citep[e.g.,][]{1997ApJ...481..601R,1998ApJ...509..661D,Carswell2002,Bergeron2002,Simcoe2004,Bergeron2004}
and pixel optical depths~\citep{1998ApJ...509..661D,2000ApJ...541L...1S,Telfer2002,2004A&A...419..811A,Pieri2004}. 

Previous pixel studies did
not attempt to convert their \OVI\ detections into oxygen abundances;
but we can compare to their recovered optical depths.
Both \citet{2000ApJ...541L...1S} and \citet{2004A&A...419..811A} claim
detection of  \OVI\ down to $\thi\approx 0.2$.   Using our combined data set binned in density (see Fig.~\ref{O_H.inv.zbin}), we obtain 1-$\sigma$ detections down to about the mean density ($\thi\approx 0.5$ at $z =2.5$).  While \OVI\ is in principle an excellent tracer of metal in very low-density gas, in practice we find that for the higher redshifts where low densities are more easily probed, \HI\ contamination is severe. Thus we are not in practice able to constrain metals in underdense gas as claimed in previous studies, in spite of a large sample and improved techniques of removing contaminants. On the high-$\thi$ side, both~\citet{2004A&A...419..811A} and~\citet{Pieri2004} exclude pixels
saturated in \HI, so cannot probe $\log \thi \gtrsim 0.5$. (This accounts,
for example, for our detection of \OVI\ in Q1422+230,
while~\citet{Pieri2004} had no detection in the same QSO).  The study
of~\citet{2000ApJ...541L...1S} did probe high $\thi$, where their
results are broadly consistent with ours. 

The studies of \citet{Carswell2002}, \citet{Simcoe2004}, ~\citet{Telfer2002},
and~\citet{Bergeron2004} did perform ionization corrections and we can compare our abundance determinations (relatively directly) to
theirs. \citet{Carswell2002} assumed to abundance of oxygen relative to carbon
to be solar and inferred metallicities for various UVB models and a
number of \OVI\ absorbers in $N_{\rm HI} \sim 10^{15}~{\rm cm}^{-2}$
systems at $z\approx 2$. They found that for relatively hard UV backgrounds,
comparable to our model Q, the ionization models yielded densities in
agreement with theoretical predictions for self-gravitating clouds
with the observed \HI\ column densities ($n_{\rm H}\sim 10^{-4}~{\rm
  cm}^{-3}$; \cite{2001ApJ...559..507S}) and metallicities 
of $10^{-3}-10^{-2}$ solar, in excellent agreement with our measurement
of [O/C]$\approx $[C/H]$\approx -2.6$ for model Q at $z=2$ and $\log \delta
= 1.5$. We note that if \citet{Carswell2002} would have allowed oxygen
to be overabundant relative to carbon, they would have found that
softer UVB models are required to obtain density estimates that agree
with theoretical expectations for gravitationally confined clouds.

\citet{Telfer2002} employed the Faint Object Spectrograph on
\textit{HST} to search for \ion{O}{5} in QSO spectra from redshift
1.6$\leq z \leq$ 2.9.  The \OVI/\ion{O}{5} ratio found by the survey
favored a UVB background similar to our Q, which they use to derive a
metallicity of $-2.2 \la $[O/H] $\la -1.3$.  While somewhat higher than our value, much of the difference may be attributable to their use of the 78th percentile in the \ion{O}{5}/\HI\  ratio. Assuming that [O/H] has a scatter at fixed density similar to that in [C/H] (see~\cite{paper2}), this would correspond to a {\em median} of $\sim 0.5\,$dex less, or $-2.7 \la $[O/H] $\la -1.8$, in fairly good agreement with our numbers.

~\citet{Bergeron2004} divide their sample into ``metal-poor absorbers"
with $N($\OVI$)/N($\HI$) < 0.25$, which they take to be predominantly
photoionized, and ``metal-rich  
absorbers" with $N($\OVI$)/N($\HI$) > 0.25$, for which they assume a
hotter phase.  For the metal-poor systems they use a ``hard" UVB and assume [O/C]$=0$ to derive a range of $-3.0
\la $ [O/H] $ \la -1.0$, 
and for the metal-rich phase they infer a
median [O/H] $\approx -0.80 $ to $-0.33$, depending upon the
assumptions regarding the ionization balance.  Combining their samples
they estimate cosmic density $\Omega_{\rm O, IGM} \approx (1.6 - 4.4)
\times 10^{-6}$, corresponding to $-2.4 \la $ [O/H] $ \la -2.0$ if
divided by the cosmic gas density $\Omega_b=0.045$.  While precise comparison is difficult, these numbers are
consistent with our corresponding estimates of [O/H]$\approx
-2.3$, or $\Omega_{\rm O,IGM}\approx 3.3 \times 10^{-6}$ using the Q
UVB. 

\citet{Simcoe2004} \textit{assume} [O/C]=0.5 and 0.0 for UVB
backgrounds comparable to our QG and Q, for $2.2 \la z \la 2.8$, so
(as for the above studies) comparing derived [O/C] values is less useful than several other points of comparison.  
First, in both backgrounds, their dependence of
[O/H] upon $\delta$ is similar to that found for [C/H] vs.\ $\delta$
in Paper II, consistent with a constant [O/C] value.  Second,
\citet{Simcoe2004} find that there is a clear jump in in the median
      [O/H] at $\delta\sim$10, while a corresponding jump is not seen in
      [C/H], similar to our results in \S \ref{sec-ioncorr}.  They
      also interpret this as possibly indicating that stronger
      absorbers are physically more complex or multiphased. Third,
      ~\citet{Simcoe2004} compute an overall contribution $\Omega_{\rm
	O, IGM} \approx 1.0 \times 10^{-6}$, using their ``hard"
      background; this would correspond to a cosmic average
      contribution of [O/H] =$ -2.6$; these numbers are $\approx
      0.3\,$dex lower than our values, but this should be regarded as
      good agreement given the number of assumptions made in each
      computation. 

Finally, the studies of \citet{1997ApJ...481..601R}\citet{1998ApJ...509..661D} are similar to ours in employing simulated spectra to attempt to match observed \OVI\ absorption and thus constrain [O/C] and [O/H].  \citet{1997ApJ...481..601R} generated simulated spectra 
from a constant-metallicity simulation and compared ionic ratios to extant data, inferring [C/H] $\approx -2.5$ and evidence for overabundance of Si and O relative to carbon. \citet{1998ApJ...509..661D} used Q1422+230 and a quasar-only Haardt \& Madau UVB model like `Q'.  The found that their data is consistent with [O/C] $\approx 0$ and [C/H]$ \approx -2.5$; these are quite consistent with our results using the Q UVB.  As an alternative interpretation, they note that a softer UVB would give high [O/C] (more characteristic of Type II supernova yields), but also lower [C/H], again consistent with our findings; however they interpret this softness as due to patchy reionization, whereas we favor its explanation by the contribution of galaxies to the UVB.

\subsection{Collisionally Ionized Gas}
\label{sec-collisional}

As discussed above, the difference between the inferred [O/H] for
$\delta\ga10$ and $\delta \la 10$ is probably due to collisionally
ionized gas, for reasons sketched in
\S~\ref{ioncorr}. Fig.~\ref{fig:ionpred} shows that for $T \gg
10^5~$K, collisional ionization dominates and the optical depth ratios become
independent of the density. For $T \ll 10^5$~K, on the other hand, the
\OVI/\CIV\ and the \OVI/\HI\ ratios both drop rapidly with increasing
density. Consequently, these ratios can be many orders of
magnitude higher in hot, dense gas than in warm, dense gas. The lower
the density, the smaller the differences become. In fact, at high
densities the \OVI\ fraction in warm gas is too small for \OVI\ to be
observable. Therefore, any \OVI\ at the redshift of very strong \HI\
absorption is likely to arise in a different phase than the associated
\HI\ and possibly even the associated \CIV. The \OVI\ phase must
either have a much lower density or a much higher temperature.

Because the fraction
of hot gas in our simulation is small at all densities, we effectively
assume the gas to be photo-ionized when we compare with synthetic
spectra and when we correct for ionization (as in
Fig.~\ref{O_H.inv.zbin}). In the latter case, we also implicitly assume
that \OVI\ and \HI\ absorption arise in the same gas phase.    

Hence, our results from \S~\ref{sec-ioncorr} suggest the existence of
a detectable amount of \OVI\ enriched hot ($T>10^{5}\,$K) gas associated
with strong \HI\ absorption. A possible physical explanation is
that such \OVI\ systems coincide with
outer regions of high $z$ galactic halos, where the effects of
galactic winds may dominate the heating process.  If the temperature
exceeds 10$^5\,$K in these regions, then this gas would contain
significant \OVI, while lacking \CIV\ and \HI. The latter two would
then arise in a cooler gas phase, which has to be fairly dense in
order to account for the strong \HI\ absorption. The high density of
the cooler phase implies that it would not produce significant
photo-ionized \OVI. 

These results and inferences are consistent with other observational studies of \OVI.
\citet{Carswell2002} and \citet{Bergeron2002} both find that the
majority of the detected absorption systems had temperatures
determined from the line widths too low for collisional ionization,
but cannot rule out higher temperatures for some absorbers.   Indeed,
the study of \citet{Simcoe2002} find their detected high-column
density \OVI\ lines associated with strong \HI\ absorbers are broad
enough to be consistent with collisional ionization. As noted above, 
\citet{Simcoe2004} find a jump in [O/H] at $\delta\sim$10
interpretable as a transition to a regime in which collisionally
ionized gas affects the abundance inferences. 

While collisionally ionized gas complicates oxygen abundance
inferences, the flip-side is that \OVI\ then provides an important
probe of hot, enriched IG gas that is difficult to detect in \CIV\
(e.g., Paper II).  Indeed, hydrodynamical simulations by
\citet{2002ApJ...578L...5T} and \citet{Dave2006} predict that a
significant portion of \CIV\ is collisionally
ionized. Hence, it would thus be very interesting
to compare such simulations, employing \OVI\ as a metallicity tracer,
with the observations analyzed here.

\section{Conclusions}
\label{sec-conc} 

We have studied the relative abundance of oxygen in the IGM by analyzing
\OVI, \CIV, \SiIV, and \HI\ pixel optical depths derived from a set of 
high-quality VLT and Keck spectra of 17 QSOs at $2.1\la z \la 3.6$, and
we have compared them to realistic, synthetic spectra drawn from a
hydrodynamical simulation to which metals have been added.  Our fiducial
model employs the ionizing background model (``QG'') taken from Haardt
\& Madau (2001) for quasars and galaxies (rescaled to reproduce the
observed mean Ly$\alpha$ absorption).  The simulation assumes a
silicon abundance as calculated in Paper III, [Si/C]$=0.77 \pm 0.05$,
and a carbon abundance as derived in  
Paper II: at a given overdensity $\delta$ and redshift $z$, [C/H] has a
lognormal probability distribution centered on
$-3.47+0.65(\log\delta-0.5)$ and of width 0.70~dex. 
The main
conclusions from this analysis are as follows:
 
\begin{itemize}

\item For 1.9$\leq z \leq$3.6, $\thi\leq 10$, and $\delta\leq$10
  (when smoothed on the scale of the \HI\ absorption, $10-10^2$
  kpc), the  fiducial simulation utilized in Papers I, II, and III consistently
  agrees with the observed $\tovi/\tciv(\tciv)$, $\tovi/\thi(\thi)$, and (to a
  lesser degree) $\tovi/\tsiiv(\tsiiv)$. Fitting $\tovi/\tciv(\tciv)$ yields a constant [O/C] = 0.66 $\pm$ 0.06, with estimated systematic errors within $\pm 0.2$ dex. Converting the observed $\tovi/\thi(\thi)$ into [O/C]$(\delta)$ using the ionization correction method of Paper II further supports these results.

\item
The relative abundances [O/C] and (especially) [O/Si] are
  sensitive to the UVB shape. We find that our fiducial (Haardt \&
  Madau 2001, quasars and galaxies) spectrum gives reasonable results
  for both, but that significantly softer or harder UVBs, such as the
  Haardt \& Madau 2001 quasar-only UVB, give results that are highly
  inconsistent with both theoretical yields and observed abundance
  ratios in other low-metallicity
  environments, and should not be considered tenable.

\item 
Our results,  both from applying the ionization correction and
  from comparing the simulations to the observations, suggest no evolution
  in [O/H] over the redshift range 1.9$\la z \la$3.6, but  a strong
  dependence on 
  $\delta$.  Both results are consistent with those found in Paper II
  for [C/H].

\item
For $\thi \geq 10$ and $\delta$ $\geq$ 10 the value of [O/C]
  (derived by comparison to the simulations) is inconsistent with that
  found at lower densities, and $\sim 0.5\,$dex higher than that
  predicted using the carbon distribution of Paper II and a
  density-independent [O/C] value. This might in principle suggest a
  density-dependent [O/C] ratio, but we favor the interpretation that
  a fraction of the high-$\delta$ \OVI\ absorbing gas is collisionally
  ionized, and that this leads to an erroneously large ionization
  correction in this regime. This interpretation is supported by our
  simulated spectra as well as by the observation that \OVI\ lines
  associated with strong \HI\ absorbers tend to be broader than those
  associated with weak \HI\ systems
  \citep{Carswell2002,Simcoe2002,Bergeron2002}. 

\end{itemize}

\acknowledgements We are grateful to Wallace Sargent, Michael Rauch and Tae-Sun Kim for providing the Keck/HIRES and VLT/UVES data used here and in Papers I-III.    We are also extremely grateful to Daniel Savin for his assistance in understanding and assessing the systematic uncertainties in recombination rates. Thanks also to Rob Wiersma for computing the expected Si/O ratio from nucleosynthetic yields taken from the literature. We thank the anonymous referee for providing comprehensive and helpful feedback that improved the manuscript. AA and CDH gratefully acknowledge support from NSF Grant
AST-0507117 and JS from Marie Curie Excellence Grant
MEXT-CT-2004-014112.

\end{document}